\documentclass[twocolumn,showpacs,amssymb,prd]{revtex4}

\usepackage{graphicx}
\usepackage{dcolumn}
\usepackage{bm}
\usepackage{epsfig}

\begin{document}
\newcommand{\gsim}{\hbox{\rlap{$^>$}$_\sim$}}
\newcommand{\lsim}{\hbox{\rlap{$^<$}$_\sim$}}


\title{Dark Matter Signals In Cosmic Rays?}

\author{Shlomo Dado} \affiliation{Department of Physics and Space Research 
Institute, Technion, Haifa 32000, Israel}

\author{Arnon Dar}
\affiliation{Department of Physics and Space Research
Institute, Technion, Haifa 32000, Israel}

\date{\today}

\begin{abstract} 

The flux of the diffuse gamma-ray background radiation (GBR) does not 
confirm that the excess in the flux of cosmic ray electrons between 
300-800 GeV, which was measured locally with the ATIC instrument in 
balloon flights over Antartica, is universal as expected from dark matter 
annihilation. Neither does the increase with energy of the fraction of 
positrons in the cosmic ray flux of electrons in the 10-100 GeV range 
that was measured by PAMELA imply a dark matter origin:  It is consistent 
with that expected from the sum of the two major sources of Galactic 
cosmic rays, non relativistic spherical ejecta and highly relativistic 
jets from supernova explosions.

\end{abstract}

\pacs{98.70.Sa, 98.70.Rz, 98.70.Vc}

\maketitle 

The spectrum of the diffuse $\gamma$ background radiation (GBR) that was 
measured by EGRET aboard the Compton Gamma Ray Observatory (CGRO) 
\cite{EGRET} showed an excess above 1 GeV in comparison with the flux 
expected from interactions of cosmic ray (CR) nuclei and electrons in the 
Galactic interstellar medium (ISM) \cite{Strong}. The origin of this GeV 
excess has been unknown.
Among its suggested origins was annihilation or 
decay of weakly interacting dark matter particles \cite{DMARev}. However, 
recent measurements with the Large Area Telescope (LAT) aboard the Fermi 
observatory have yielded preliminary results \cite{LAT} which 
do not show 
a GeV excess at small Galactic latitudes and agree with the flux expected 
from CR interactions in the Galactic ISM. Moreover, by comparing the 
spectra of gamma-rays around GeV from nearby Galactic pulsars, which were 
measured by EGRET and LAT, the Fermi collaboration confirmed \cite{LAT}
previous conclusions \cite{Stecker} that the 
origin of the EGRET GeV excess is instrumental and not a dark 
matter annihilation/decay signal. In this letter we show that an absence 
of a GeV excess in the GBR also challenges the reported excess in the flux 
of CR electrons at energies between 300-800 GeV that was measured with the 
Advanced Thin Ionization Calorimeter (ATIC) aboard balloon flights over 
Antartica \cite{ATIC} and was also suggested to arise from annihilation of 
dark matter particles such as Kaluza-Klein particles with a mass of about 
620 ${\rm GeV/c^2}$ \cite{DMKZ}). We also demonstrate that 
the reported 
increase in the fraction of positrons in the flux of CR electrons above 10 
GeV, which was measured with the satellite-borne Payload for Antimatter 
Exploration and Light Nuclei Astrophysics (PAMELA) \cite{PAMELA}, 
and was interpreted as a dark matter signal, 
is that expected from the main source of Galactic cosmic rays -
the non-relativistic shells and the highly relativistic 
jets ejected in supernova explosions.

Practically all CR acceleration mechanisms invoke an ionized medium that 
is swept by a moving magnetic field, such as would be carried by the 
rarefied plasma in a supernova shell or a plasmoid of a jet 
\cite{DP},\cite{DD}. The swept-in 
ionized particles, which enter the plasma with  
$\gamma\!=\! E/m\,c^2$ equal to the bulk motion Lorentz factor
of the plasma, are isotropized and Fermi accelerated 
by its turbulent magnetic field.  The deceleration of 
the jets/ejecta by the swept-in particles and their isotropic emission in 
the jet/ejecta rest frame determine their spectrum. To the extent that 
particle-specific losses (such as synchrotron radiation) can be neglected, 
all source fluxes roughly have the same power-law 
spectrum $dF_s/dE\!\sim\! E^{\!-\!\beta_s}$ with $\beta_s\!\approx\! 
2.2$ \cite{DD}. This injected spectrum of Galactic CR nuclei is modulated 
mainly by their accumulation during their residence time in the Galaxy, 
$\tau_{gal}(E)$, before they escape into the intergalactic space by 
diffusion in the Galactic magnetic field. Observations of the nuclear 
abundances of secondary CRs produced in the interstellar medium 
(ISM) as functions of energy indicate that $\tau_{gal}(E)\!\propto\! 
E^{\!-\!0.5\!\pm\! 0.1}$ \cite{Swordy}. Thus, in a steady state, the 
observed Galactic flux of CR nuclei below the knee is predicted to have a 
spectrum $dF/dE\!\propto\! \tau_{gal}\, dF_s/dE\!\propto\!  
E^{\!-\!2.7\!\pm\! 0.1}$ in good agreement with observations, while CRs
which escape into the intergalactic space have the source spectrum 
$dF/dE\!\sim\! E^{\!-\!2.2}$ \cite{DD}.

High energy CR electrons can also be produced by the interaction of CR 
nuclei with matter and radiation, mainly through production and decay of 
charged pions: $\pi\!\rightarrow \! \mu\!+\!\nu_\mu,$ $\mu \!\rightarrow\! 
e\!+\!\nu_e\!+\!\nu_\mu$.  Due to Fynman scaling, they also have a 
power-law source spectrum with a spectral index $\beta_s\!=\!2.2$, if 
produced in/near the source of CR nuclei or in the inter galactic medium 
(IGM), or 
$\beta_s\!=\!2.7$ if produced by CR interactions in the Galactic ISM 
\cite{Dar83}. But,
unlike CR nuclei, CR electrons in the ISM above a few GeV lose their 
energy by inverse Compton scattering (ICS) of starlight and cosmic 
microwave background radiation (MBR) and by synchrotron radiation in the 
ISM magnetic field, in a time shorter than the Galactic confinement time 
\cite{DD2001},\cite{DD}:
\begin{equation}
\tau_{cool}(E_e)={3\,(m_e\, c^2)^2 \over 4\,\sigma_{_T}\, c\, 
E_e\,[\rho_\gamma+\rho_{_B}]}\approx {2.85\times 10^8\,{\rm y}\over 
(E_e/{\rm GeV})}\,,
\label{tcool}
\end{equation}
where $\sigma_{_T}\!\approx\!0.665\times 10^{\!-\!24}\,{\rm cm^{\!-\!2}}$ 
is the Thomson cross 
section, $\rho_\gamma\!=\!0.47\,{\rm eV\, cm^{\!-\!3}}$ is the 
mean energy density of starlight
and MBR, and $\rho_{_B}\!=\! B^2/8\, \pi \!\approx\! 
0.62\, {\rm 
eV\,cm^{\!-\!3}}$ is the mean energy density of the magnetic field 
($B\!\sim\! 
5\, \mu{\rm Gauss}$) in the Galactic disk. This implies a spectrum of high 
energy CR 
electrons, $dF_e/dE\!\propto\!  E^{\!-\!\beta_s\!-\!1}\!=\!E^{\!-\!3.2}$ 
\cite{DD2001}, \cite{DD}. Radio and X-ray observations of synchrotron 
radiation emitted by CR electrons accelerated by the major cosmic accelerators such 
as supernova remnants, gamma ray bursts, microquasars and active galactic 
nuclei, confirm this predicted universal spectrum of high-energy CR 
electrons. Indeed, the energy spectrum of CR electrons with $E_e\!>\!5$ 
GeV that was measured directly near Earth 
\cite{HEAT},\cite{AMS},\cite{Kobayashi},\cite{BETS} is well described by:
\begin{equation}
{dF_e\over dE}=(1.5 \pm 0.5)\times 10^5\, 
\left [E\over {\rm MeV} \right]^{-3.2\pm 0.1}{1\over {\rm 
cm^2\, s\, sr\, MeV}}. 
\label{CRe}
\end{equation} 
Inverse Compton scattering of MBR photons and 
stellar light by these electrons give rise to a diffuse GBR with 
a power-law spectrum $dF_\gamma/ dE\!\sim\! E^{\!-\!\beta_\gamma}$ and a 
power-law index $\beta_\gamma\!=\!(\beta_e\!+\!1)/2 \approx 2.1\pm 0.05\, 
.$  It dominates the diffuse GBR at large Galactic latitudes and 
contributes significantly to the diffuse GBR at small Galactic latitudes.

\begin{figure}[]
\centering
\epsfig{file=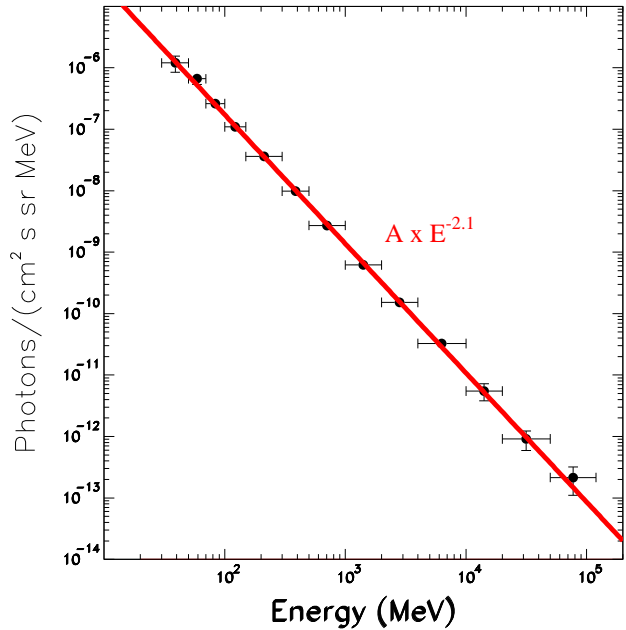,width=9.2cm}
\vspace*{-20pt}
\caption{The GBR spectrum, inferred by EGRET \cite{Sreek}.
The line is the best power-law fit.}
\label{GBRspectrum} 
\end{figure}

The intensity and spectrum of the diffuse GBR was measured by EGRET aboard 
the Compton Gamma Ray Observatory (CGRO). An extragalactic GBR was 
inferred ~\cite{Sreek} from the extrapolation of these measurements
in directions away from the Galactic disk and center to zero column density, 
in order to eliminate the Galactic contributions of bremsstrahlung from 
CR electrons, and $\pi^0$ production by CR nuclei in the ISM.
Fig.~\ref{GBRspectrum} shows that this GBR flux in the 30~MeV-${\rm 
120~GeV}$ energy range is well described by a single power-law:
\begin{equation}
\!{dF_\gamma\over dE}\!\simeq\! (2.7\pm 0.1)\times 10^{-3} 
 \left[{E\over {\rm MeV}} 
\right]^{-2.1\pm 0.03}\! {1\over {\rm~cm^{2}~s~sr~MeV}}\,.
\label{photons}
\end{equation}
The spectral index of the GBR inferred by EGRET is the same, $2.1\pm 
0.03,$ in all sky directions away from the Galactic disk \cite{Sreek}. But 
the data show a 
significant deviation from isotropy, clearly correlated with the structure 
of the Galaxy and our position relative to its center \cite{DD2001}. This 
advocates a large Galactic contribution to the EGRET GBR arising from ICS 
of MBR photons and starlight by Galactic CR electrons \cite{DD2001}. The 
uniformity of the spectral index of the diffuse GBR over the whole sky, 
despite a large Galactic contribution correlated with the structure of the 
Galactic CR halo and our position relative to its center, suggests a CR 
origin of both the Galactic and extragalactic diffuse 
GBR \cite{DD},\cite{D2007}.

Below $\!\sim\!100$~GeV the spectrum of Galactic CR electrons measured by 
ATIC agrees with that measured in other experiments  and is well 
represented by Eq.~(\ref{CRe}). Above $\!\sim\!100$~GeV the ATIC 
results show an excess in flux of CR electrons that peaks around 
$\sim$650~GeV and drops rapidly to zero around 800~GeV.  Consider  
ICS of MBR photons whose
present temperature and mean energy are 
${T_0\!=\!2.725}$ K and $\epsilon\!\approx\! 2.7\, k\,T_0\!\approx\! 
0.635\,{\rm meV}$ \cite{MBR}. The mean energy of these 
upscattered photons by
CR electrons with $E_e\!\sim\!650$~GeV is: 
\begin{equation} 
{E_\gamma(\epsilon) \approx {4\over 3}\,
\left({E_e\over m_e\, c^2}\right)^2\epsilon}\approx 1.37~{\rm GeV}\, .
 \label{loss} 
\end{equation} 
This relation holds for the observed $E_\gamma$ independent of the
redshift where the ICS took place
because the blueshift of the MBR temperature   
compensates the redshift of the observed 
energy of the scattered photons. Inverse Compton scattering 
of stellar photons ($\epsilon\!\sim\!1$~eV) by 650 GeV electrons
boosts the energy of the stellar photons to $\sim$2~TeV, but this process
is strongly suppressed by the energy dependence of the 
Klein-Nishina cross section.

\begin{figure}[]
\centering
\epsfig{file=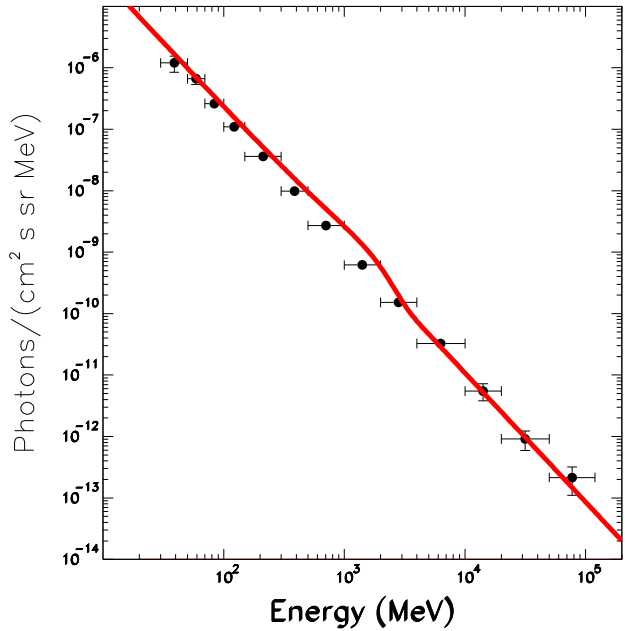,width=9.2cm}
\vspace*{-20pt}
\caption{Comparison between the spectrum of the GBR which was measured by 
EGRET \cite{Sreek} and that produced by ICS of MBR photons by  
CR electrons with a power-law spectrum with  
an index $\beta_e\!=\!3.2$ plus  an 
excess such as that measured by ATIC \cite{ATIC} between 300-800 
GeV.}
\label{GBRATIC1}
\end{figure}

\begin{figure}[]
\centering
\epsfig{file=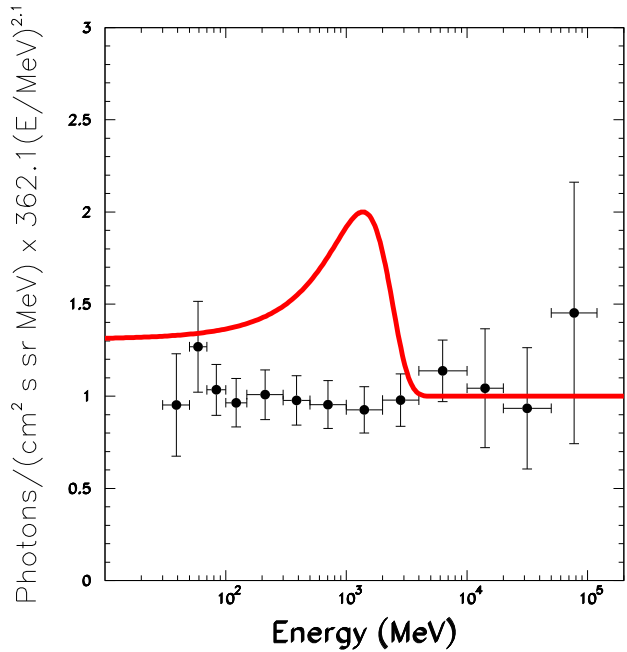,width=9.2cm}
\vspace*{-20pt}
\caption{Comparison between the GBR spectrum measured by EGRET 
\cite{Sreek} and a GBR spectrum produced by ICS of MBR photons by a 
universal CR electrons power-law spectrum plus an 
excess such as that measured by ATIC \cite{ATIC} between  300-800 GeV.
Both spectra were divided by the best fitted power-law to the EGRET 
GBR spectrum \cite{Sreek}.}
\label{GBRATIC2}
\end{figure}

The GBR flux produced by ICS of MBR photons at Galactic 
latitude $b$ and longitude $l$ is given to a good approximation by 
[DD2001]:
\begin{equation}
{dF_\gamma\over dE_\gamma}\simeq N_e(b,l)\, \sigma_T\, 
\left[{dE_e\over dE_\gamma}\, {dF_e\over dE_e}\right]_{E_e=E_{eff}}\,,
\label{ICSflux}
\end{equation}
where $N_e(b,l)$ is the column density of CR electrons at latitude $b$ 
and longitude $l$ and $E_{eff}\!=\!m_e c^2\sqrt{3\, 
E_\gamma/4\,\epsilon}.$
The ATIC flux excess between 500-800 GeV and the smooth CR electron flux 
without it are comparable. Thus, if the 
ATIC excess between 500-800~GeV is `universal' and not a 
local excess, such as that 
expected from dark matter annihilation in the entire universe, 
it should 
have produced an excess in the GBR between 0.8-2~GeV ( i.e., a `GeV 
excess') comparable with the contribution from ICS of MBR photons by a CR 
electron flux without the ATIC excess. 
This `GeV excess' in the GBR should 
be observable mainly at large Galactic latitudes where ICS 
scattering of MBR photons dominates over photons from the decay of $\pi^0$ 
produced in collisions 
of high energy CR nuclei in the ISM. An estimate of the expected GBR at 
large Galactic latitudes, which is based on Eq.~(\ref{ICSflux}) and the 
universality of the ATIC excess, is shown in  Figures~\ref{GBRATIC1}
and \ref{GBRATIC2}.
These Figures show that the
EGRET GBR does not support a `universal' 
ATIC excess ($\chi^2/dof\!=\!14.6 $ for an ATIC-like flux compared 
to 
$\chi^2/dof\!=\!0.27 $ for the power-law fit in 
Figure~\ref{GBRspectrum}.)

Very bright local point sources of CR electrons with $E_e\!\sim\! 650$ GeV 
could produce the ATIC excess \cite{Pulsars}. 
But, such sources must also be extremely bright in GeV and TeV 
$\gamma$-rays due to ICS of the MBR and stellar light near the sources. 
Ultra-bright nearby steady sources of $\gamma$ rays were not observed 
either with EGRET aboard the CGRO satellite and LAT aboard the Fermi 
satellite in the GeV range or with the H.E.S.S., MAGIC and VERITAS air 
shower Cherenkov telescopes in the TeV range. Moreover, steady 
or transient bright sources of Galactic CRs such as gamma 
ray bursts that are beamed away from Earth \cite{DP}, which could have 
produced a local CR excess like the ATIC excess, 
would have also produced such a universal excess that is ruled 
out by the observed GBR.

\begin{figure}[]
\centering
\epsfig{file=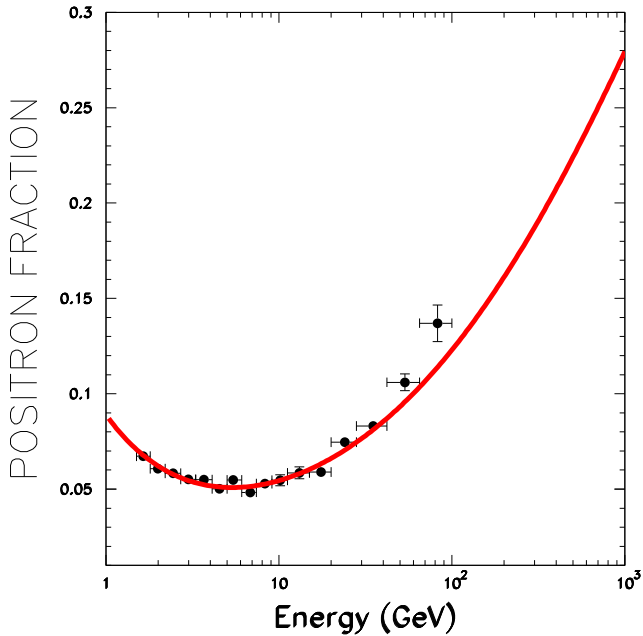,width=9.2cm}
\vspace*{-20pt}
\caption{Comparison between
Eq.~(\ref{posfrac}), which follows from CR acceleration 
by both, non relativistic spherical ejecta and highly relativistic
jets from supernova explosions \cite{DP}, \cite{DD}, and 
the PAMELA positron fraction \cite{PAMELA}.} 
\label{pefrac}
\end{figure}

The increasing positron fraction in the 10-100 GeV range which was 
measured by PAMELA 
is at odds with the expected decrease from the decay of $\pi'{\rm s}$ and 
$K's$ produced in the collisions of the primary CR nuclei with ISM nuclei 
\cite{fraction}. The observed increase may result from misidentified 
hadrons by PAMELA (whose fraction may increase with energy due to a more 
rapid decline with energy of the positron flux than the flux of protons 
and secondary hadrons \cite{misidentified}). However, the flat positron 
fraction between 5-50 GeV that was measured by HEAT \cite{HEAT} with the 
record hadron rejection ($\geq 10^{\!-\!5}$) is consistent within errors 
with the PAMELA result but is also inconsistent with the slow decrease 
with increasing CR energy from secondary meson production  
in the ISM calculated from a Leaky Box 
Model \cite{fraction}. It suggests that another positron source begins to 
dominate the CR positron flux around 10 GeV.  Positrons of 10-100 GeV 
energy Compton upscatter starlight ($\epsilon\!\sim\!1$~eV) to $E_\gamma 
\!\sim\! 0.5-50$ GeV. But, unlike the ATIC excess, the small positron 
fraction measured by PAMELA (Figure~\ref{pefrac}) have too small a 
signature which can be resolved from the GBR with current instrumentation 
aboard gamma-ray satellites.

The increaing positron fraction measured by PAMELA could be either  
local or global. The sources which can enhance 
the positron fraction locally could be: (i) a local
environment with a  higher density than that of the ISM;
(ii) nearby luminous 
sources  of CR nuclei (such as supernova remnants) which can
enhance local production of CR electrons and positrons in the ISM;
(iii) nearby astrophysical point sources of CR positrons    
such as pulsars which are  believed to be producers  of high energy 
$e^+e^-$ pairs \cite{Pulsars}. Global sources include
gamma ray bursters whose jets can transport ballisticly primary CRs
to Galactic distances \cite{DP},\cite{DD} and 
annihilation/decay of dark matter particles \cite{DMKZ}.  

Consider the combination of primary electrons produced 
by the CR sources and the secondary
CR electrons produced in the ISM.  At low energies, the observed 
positron fraction near minimum solar activity decreases like $\!\sim\! 
E^{-0.8}$ because of the effects of the geomagnetic field, which are not 
fully understood, and because multi-meson production becomes more 
symmetric in charge with increasing energy. At high energies the electrons 
and positrons which are produced in the CR sources are injected into the 
ISM with $\beta_s\!=\!2.2$ and a positron fraction $\!\sim\!0.7\,,$ and 
cool to a power-law $\!\propto\! E^{\!-\!3.2},$ while the secondary 
electrons that are injected with a spectrum $\!\propto\! E^{\!-\!2.7}$ 
cool to a steady state spectrum $\!\propto\! E^{\!-\!3.7}$.
Thus, the positron fraction can be described effectively by: 
\begin{equation}
{e^+ \over e^+ + e^-}\approx {E^{-4.5}+a\, E^{-3.2} \over 
                         b\,E^{-3.7}+(a/0.7)\, E^{-3.2}} \, ,
\label{posfrac}
\end{equation}    
which interpolates between the expected decrease at low-energy  
and the increase towards $\!\sim\!0.7$ at higher energies. 
In Figure~\ref{pefrac} we compare Eq.~(\ref{posfrac}) and the PAMELA
positron fraction \cite{PAMELA}. The 
values of $a$ and $b$, which depend on the phase of the solar 
cycle and on many other not well known  astrophysical parameters,
were  treated as adjustable parameters. For $E$ in GeV units, their best 
fitted values are, $a\!=\!0.19$ and $b\!=\!13.1\,$. 
Eq.~(\ref{posfrac}) seems to fit well 
the positron fraction measured by PAMELA without invoking 
positron production by dark matter annihilation. It is  
consistent with the fact that the antiproton flux measured
by PAMELA \cite{PAMELA} and in other experiments agrees
with that expected from CR production of these antiprotons in the ISM.  

In conclusion, the measured GBR rules out a universal source, such as dark 
matter annihilation, as the origin of the ATIC excess. 
Positron production in/near source plus production in the ISM
can explain the behaviour of the positron fraction measured by PAMELA.


\begin{thebibliography}{999}

\bibitem{EGRET}%
S. D. Hunter {\it et al.}  {\it Astrophys. J.} {\bf 481}, 205
(1997).

\bibitem{Strong}%
A. Strong \& I. V.  Moskalenko, {\it Astrophys. J.} {\bf 509}, 212 (1998);

\bibitem{DMARev}%
See, e.g., W. de Boer, C. Sander, V. Zhukov, A. V. Gladyshev \& 
D. I. Kazakov, {\it Phys. Rev. Lett.} {\bf
95}, 209001 (2005). For a general review of possible dark matter signals,
see, e.g., L. Bergstrom, {\it Rept. Prog. Phys.} {\bf 63}, 793 (2000).

\bibitem{LAT}%
G. Johannesson (Fermi LAT collaboration), {\it Proc. XLIV Rencontre De
Moriond,  High Energy Phenomena in the Universe}, 
(La Thuile Italy, Feb. 2009) to be published.

\bibitem{Stecker}%
F. W. Stecker, S. D. Hunter \& D. A. Kniffen,
{\it Astropart. Phys.} {\bf 29}, 25 (2008).

\bibitem{ATIC}%
J. Chang {\it  et al}. {\it Nature}, {\bf 452}, 362 (2008).

\bibitem{DMKZ}%
H. C. Cheng, J. L. Feng \& K. T. Matchev, {\it Phys. Rev. 
Lett.} {\bf 89}, 211301 (2002); D. Hooper \& K. Zurek, arXiv:0902.0593 
(2009) and references therein.

\bibitem{PAMELA}%
A. Morselli \& I. V. Moskalenko,  arXiv:0811.3526 (2008).

\bibitem{DP}%
A. Dar \& R. Plaga, {\it Astron. Astrophys.} {\bf 349}, 259 (1999).

\bibitem{DD}%
A. Dar \& A. De R\'ujula, {\it Phys. Rep.} {\bf 466}, 179, (2008).

\bibitem{Swordy}%
S. P. Swordy {\it et al.},  {\it Astrophys. J.} {\bf 330}, 625 (1990).

\bibitem{Dar83}%
A. Dar {\it Phys. Rev. Lett.} {\bf 51}, 227 (1983).

\bibitem{DD2001}%
A. Dar \&  A. De R\'ujula, Mon. Not. Roy. Astr. Soc. {\bf 323},
391 (2001);
A. Dar, A. De R\'ujula \& N. Antoniou, {\it Proc. Vulcano Workshop 1999}
(eds. F. Giovanelli \& G. Mannocchi) p. 51,
Italian Physical Society, Bologna, Italy.

\bibitem{D2007}%
A. Dar, {\it Nucl. Phys. S.} {\bf 165}, 103 (2007).

\bibitem{HEAT}%
S. W. Barwick  {\it et al}. {\it Astrophys. J.} {\bf 498}, 779 (1998);
M. A. DuVernois {\it et al}. {\it Astrophys. J.} {\bf 559}, 296 (2001).

\bibitem{AMS}%
M. Aguilar {\it et al.}, {\it Phys. Rep.} {\bf 366}, 331 (2002)
and references therein.

\bibitem{Kobayashi}%
T. Kobayashi {\it et al}. {\it Astrophys. J.} {\bf 601}, 340 (2004).

\bibitem{BETS}%
S. Torii {\it et al}. {\it Astrophys. J.} {bf 559}, 973 (2001);
S. Torii {\it et al}. arXiv:0809.0760 (2008).


\bibitem{Sreek}%
P. Sreekumar {\it et al.,} {\it Astrophys. J.} {\bf 494}, 523 (1998).


\bibitem{MBR}%
J. C. Mather {\it et al.},  {\it Astrophys. J.} {\bf 432}, L15, (1993);
D. J. Fixsen {\it et al.}, {\it Astrophys. J.} {\bf 473}, 576 (1996).

\bibitem{Pulsars}%
S. Profumo, arXiv:0812.4457 (2008) and references therein. 

\bibitem{misidentified}%
M. Schubnel, {\it Proc. XLIV  Rencontre De
Moriond,  High Energy Phenomena in the Universe},
(La Thuile Italy, Feb. 2009) to be published.

\bibitem{fraction}%
I. V. Moskalenko \& A. W. Strong, {\it Astrophys. J.} {\bf 493}, 694 
(1998).

\end{thebibliography}
\end{document}